\documentstyle[psfig]{aa}
\newcommand{\ark}{\hbox{Ark~564} }
\newcommand{\arkp}{\hbox{Ark~564.} }
\newcommand{\etal}{et al. }
\def\simlt{\lower.5ex\hbox{\ltsima}}            % < over ~
\def\simgt{\lower.5ex\hbox{\gtsima}}            % > over ~

\def\flux{erg\,cm$^{-2}\,s^{-1}$}
\begin{document}
\thesaurus{01(11.01.2; 11.19.1; 11.09.1; 11.16.01; 13.25.2) }
\title{Optical and X-ray monitoring of the NLS1 Galaxy Ark~564}
\author{I.E. Papadakis\inst{1} \and W. Brinkmann\inst{2} 
 \and H. Negoro \inst{3} \and E. Detsis\inst{1}
 \and I. Papamastorakis\inst{1,4} \and M. Gliozzi\inst{2} } 
\offprints{I. E. Papadakis;  e-mail: jhep@physics.uoc.gr}
\institute{Physics Department, University of Crete, P.O. Box 2208,
   71003 Heraklion, Greece 
\and Max--Planck--Institut f\"ur extraterrestrische Physik,
   Giessenbachstrasse, D-85740 Garching, Germany
\and Cosmic Radiation Laboratory, RIKEN, 2-1 Hirosawa, Wako-shi,
   Saitama 351-01, Japan
\and Foundation-for Research and Technology-Hellas, P.O.Box 1527, 711 10
heraklion, Crete, Greece}
\date{Received  ?; accepted ?}
\maketitle
 
\begin{abstract}

The Narrow-line Seyfert galaxy \ark was monitored between June 04, 2000
and June 24, 2000 from Skinakas Observatory. The light curves obtained in
$B, V, R$ and $I$ bands show night-to-night variations of $\Delta m \sim
\pm 0.1$ mag. Comparison with X-ray data from a simultaneous ASCA
long-look observation shows that the variations in X-rays are much more
pronounced than in the optical band. A cross correlation analysis yields a
time lag of $\tau \sim 0.8$ days between X-rays and optical light (with
the X-rays leading the optical), however with low statistical
significance. We discuss possible implications of our results for
physical models involving the reprocessing of the primary X-rays by the
accretion disc.

\end{abstract}

   
\section{INTRODUCTION}
\smallskip
Narrow-line Seyfert~1 (NLS1) galaxies are a peculiar group of AGN
characterized by their optical line properties: H$\beta$ FWHM does not
exceed 2000 km~s$^{-1}$, the [O~III]$\lambda$5007 to H$\beta$ ratio is
less than 3, and the UV$-$visual spectrum is usually rich in
high-ionization lines and strong Fe~II emission multiplets (Osterbrock \&
Pogge 1985). In hard X-ray studies NLS1 galaxies comprise less than 10\%
of the Seyfert galaxies, however, from the ROSAT All-Sky Survey it
 became clear that about half of the AGN in soft X-ray selected samples 
are NLS1 galaxies (Grupe 1996, Hasinger 1997).
 
Boller \etal (1996) found from ROSAT observations that the soft X-ray
spectra of NLS1 galaxies are systematically steeper than those of broad
line Seyfert galaxies and that an anti-correlation exists between the
X-ray photon index and the FWHM of the H$\beta$ line which provides strong
evidence for a physical link between the intrinsic properties of the
continuum emission and the dynamics of the broad line region. They further
discovered that NLS1 galaxies frequently show rapid short time scale X-ray
variability which can be interpreted as evidence for small black hole
masses in these objects. Further examples of extreme X-ray variability
were found in monitoring observations of IRAS~13324-3809 (Boller \etal
1997) and PHL~1092 (Brandt \etal 1999).
 
An ASCA observation of RE~1034+39 revealed a very strong soft X-ray excess
component dominating the spectrum below $\sim$ 2~keV and a very steep hard
X-ray spectrum with photon index $\Gamma \sim 2.6$ (Pounds \etal 1995).
This prompted Pounds \etal to propose that NLS1 galaxies are the
supermassive analogs of Galactic black hole candidates in the soft state.
This analogy supports the idea that the properties of NLS1 galaxies
originate from accretion near the Eddington rate onto a central black hole
of moderate mass.
  
Recent detailed spectral and variability X-ray studies of NLS1 galaxies
show that these objects exhibit peculiar spectral features which have not
been observed in Seyfert~1 galaxies with broad optical lines (for a
comprehensive overview see Leighly 1999 a,b). The high accretion rate,
generally thought to be the determining parameter of the systems, would
lead to an increasing ionizing flux and the observed spectral features can
be explained as arising from an ionized disc (Pounds \& Vaughan 2000).
   
Ark~564 is the X-ray brightest NLS1 galaxy with a 2$-$10~keV flux of
$\sim 2\times10^{-11}$ \flux (Vaughan \etal 1999).
Its X-ray properties are typical of NLS1 galaxies. It shows no neutral
absorption in the rest frame of the galaxy, nor warm absorption due to
ionized material. Its soft X-ray continuum has a steep slope 
($\Gamma > 3$, Brandt \etal 1994), due to the presence of strong soft
excess emission and the hard X-ray band energy
spectrum is steep ($\Gamma\sim2.6$, Leighly 1999b).

The soft and hard X-ray flux of \ark is variable both on short and
long time scales (Brandt \etal 1994, Leighly 1999a).
In an ASCA observation of $\sim 1$ day the
$0.5-2$ keV light curve showed an overall variation by a factor $\sim 3$.
Its excess variance was found to be similar to that of broad line Seyfert~1
galaxies but the variability is stronger in the hard energy band
while broad line Seyferts vary more at lower energies (Edelson 2000).
Interestingly, the source also showed spectral variations which were
uncorrelated with the source flux (Leighly 1999a).

In view of the strong X-ray variability displayed by NLS1 galaxies it
appears surprising that in the optical band the objects seem to be less
variable than broad line Seyferts (Giannuzzo \etal 1998). In an attempt
to evaluate the line - to - continuum time lag Shemmer \& Netzer (1999) 
found no significant variability from \ark on long or short time scales.
 
A comparison of the variations in different energy bands can provide valuable
information for the understanding of the geometry and the nature of  an 
AGN. In particular, time lags  between different wave bands have been used
to discriminate between primary and reprocessed emission and to 
obtain bounds on the size of the emission region. However, so far only
relatively few X-ray - and UV/optical monitoring programs have been
carried out for only a few Seyfert galaxies (see Maoz \etal 2000 and
references therein).
   
We conducted an optical monitoring campaign on \ark at the Skinakas
observatory in Crete simultaneously with the ASCA AO8 long-look
observation of the source. In the next section we will give the details of
the observations, then describe the cross-correlation analysis between the
optical and X-ray data and finally discuss the results.

\section{Observations and Data Reductions}

\subsection{The optical observations}

We conducted an optical monitoring campaign of \ark from June 04, 2000 to
June 24, 2000 at Skinakas Observatory in Crete, Greece. The telescope is
an 1.3 m, f/7.7 Ritchey-Cretien. The observations were carried out through
standard Johnson $B,V$ and Cousins $R,I$ filters and the CCD used was a
$1024\times1024$ Tektronix chip with 24$\mu$m pixels (corresponding to
0.5$^{\prime\prime}$ on sky). The exposure time was 60, 30, 20 and 10 sec
for the $B,V,R$ and $I$ filters respectively. During this 21 day period we
obtained data for 17 nights. The object was observed between two and six
times each night; in total, there are 60 frames in each filter.The seeing
during the observing run was between $\sim 1^{\prime\prime}$ and $\sim
3^{\prime\prime}$. Standard image processing (bias subtraction and flat
fielding using twilight-sky exposures) was applied to all images.

\begin{figure}
\begin{center}
\psfig{figure=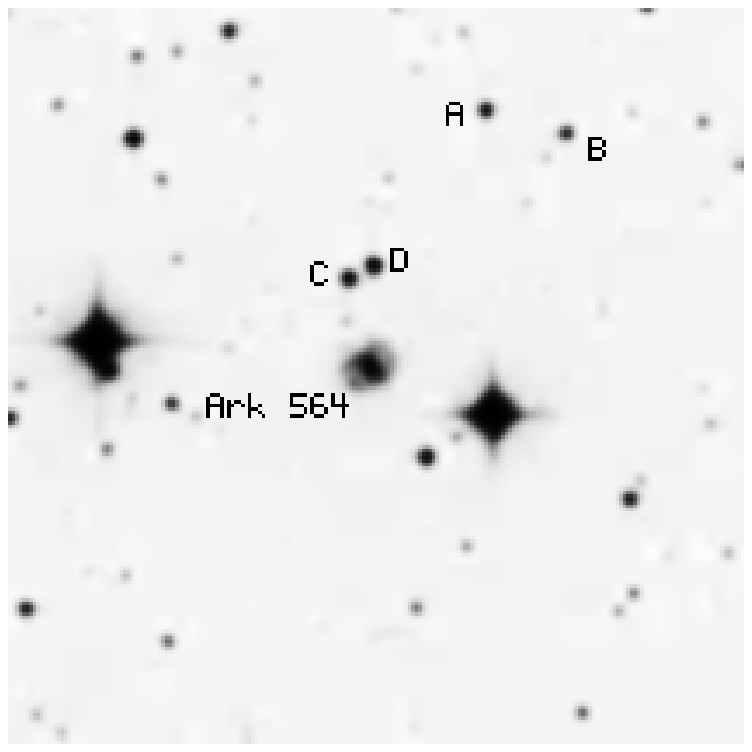,height=7.5truecm,width=8.0truecm,angle=0,%
 bbllx=190pt,bblly=290pt,bburx=410pt,bbury=505pt}
\caption[]{Finding chart ($6^{\prime}\times6^{\prime})$ of \arkp North is
up, East is left.}
\end{center}
\end{figure}

We did not find any published comparison star sequence in the field of
\arkp We therefore chose four unsaturated stars, close to the galaxy (see
Figure~1) and carried out aperture photometry by integrating counts
within circular aperture centered on each star.  The aperture had a
radius equal to 4 times the seeing full-width at half-maximum (FWHM) of
each frame in order to include all the light from the stars. Their
instrumental magnitudes were transformed to the standard system through
observations of standard stars from Landolt (1992) during the last four
days of the observing run. Table~1 lists the magnitudes of the reference
stars on the field of \ark; quoted errors include both the uncertainty in
the determination of the instrumental magnitudes (based on photon
statistics) and the uncertainty associated with the transformation to the
standard system.

\begin{table}
\begin{center}
\caption{Reference stars in the field of \ark}  
\begin{tabular}{lcccc}
\hline
   & $B$ & $V$ & $R$ & $I$ \\
\hline
A  & $16.01\pm0.03$ & $14.77\pm0.02$ & $14.47\pm0.02$ & $14.73\pm0.02$ \\
B  & $16.34\pm0.03$ & $15.09\pm0.02$ & $14.78\pm0.02$ & $15.03\pm0.02$ \\
C  & $15.43\pm0.03$ & $14.11\pm0.02$ & $13.76\pm0.02$ & $13.95\pm0.02$ \\ 
D  & $14.81\pm0.03$ & $13.65\pm0.02$ & $13.37\pm0.02$ & $13.65\pm0.02$ \\
\hline
\end{tabular}
\end{center}
\end{table}

For the AGN nucleus aperture photometry was carried out using the same
size aperture as for the comparison stars. Since the host galaxy is much
less affected by seeing variations, its contribution to the counts within
an aperture depends on the aperture's radius. We corrected for this effect
by plotting the instrumental magnitudes vs FWHM for each band. We then
fitted the data with a straight line and used the best-fitting parameters
to scale the host galaxy contribution at each aperture to its contribution
in the largest aperture.

To assure that the comparison stars are constant we performed relative
photometry by calculating the difference between a star's instrumental
magnitude in a given epoch and its magnitude in a certain frame, the
``reference frame". For each of the reference stars, these differences
were averaged using the remaining three stars as comparison stars. The
standard deviation of the resulting light curves was 0.05, 0.03, 0.02 and
0.02 mag in the $B,V,R$ and $I$ band, respectively (see Fig.~2). These
values provide an estimate of the typical error on the magnitude
measurement of \arkp Since during some nights the light curves show
scatter larger than average, we used the standard deviation of the
reference stars during each night as a measurement of the uncertainty of
the object's magnitude at the same night.

Finally, the redenning correction in the direction of \ark was taken
$A_{V}=1.24$ mag. It was derived using $E(B-V)=0.4$ (Walter and Fink 1993)
and the conversion formula of Cardelli et al. (1989) with $R=3.1$. From
the curve of Cardelli et al. we also derived the extinction (in
magnitudes) in the other filters as well: $A_{B}=1.66, A_{R}=0.93$ and
$A_{I}=0.59$.

\subsection{The X-ray data}

\begin{figure}
\psfig{figure=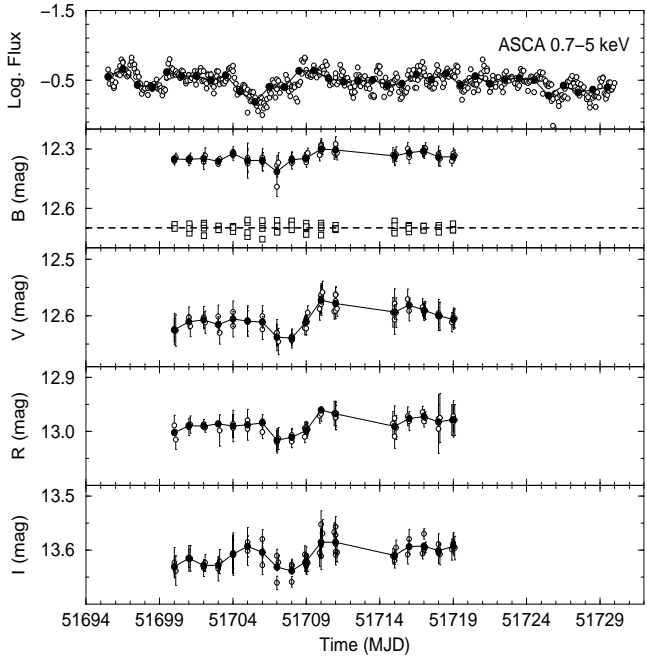,height=8.5truecm,width=8.5truecm,angle=0,%
 bbllx=10pt,bblly=60pt,bburx=540pt,bbury=550pt}
\caption[]{ Plot of the ASCA SIS $0.7-5$ keV (in arbitrary units), $B,V,R$
and $I$ optical band light curves. Open circles show the ASCA light curve
binned in 5640 s ($\sim$ orbital period) and the individual optical
measurements each night. Filled circles show the ASCA light curve binned
in one-day and the mean optical magnitudes per night. In the second from
the top plot we also show (open squares) the B relative magnitude 
of the reference stars using the other three stars as comparisons.
The light curve is shifted from mag=0 to mag=12.7 for comparison reasons.}
\end{figure}

ASCA observed \ark from 2000 June 01 11:51:27 to 2000 July 05 23:57:54.  
We extracted the X-ray data from the public archive at ISAS and applied
standard criteria for the data analysis in order to create light curves
for all four instruments. Source counts were accumulated from a $\sim
8\arcmin \times 8\arcmin$ rectangle in detector coordinates to maximize
the X-ray signal for the SIS0/1. For the GIS2 and GIS3 we used a 6 arcmin
radius centered on the source. The background was estimated from a source
free region on the same chips from the same observation. In the case of
the GIS detectors we chose a region at an off-axis distance of $\sim$
11\arcmin ~as the source was very bright. In general, the background level
was less than 3\% of the source count rate.
 
During the analysis it became apparent that all four instruments suffered
from an unexpected light leak. This is probably due to an expansion of the
atmosphere caused by recent solar activity and to the low satellite orbit
during the observation. We investigated this light leak effect using
energy spectra, low energy light curves, and DFE curves for each of the
individual, about three day long, data sets (each FRF file). We found that
in the data screening the BR\_EARTH criterion should not only be applied
to the SIS data but to the GIS data as well. Furthermore, the BR\_EARTH
criterion for the SIS was changed from $20^{\circ}$ to 35$^{\circ}$
and for the GIS from 12.5$^{\circ}$ to 17.5$^{\circ}$.
 
\section{Data Analysis}

\subsection{Variability Amplitudes}

The de-reddened $B,V,R$ and $I$ light curves (in magnitudes) together with
the SIS0/1 $0.7-5$ keV light curve are shown in Fig.~2.  Large amplitude
variations are seen in the X-ray band on all time scales while in the
optical band we observe variations of a much smaller amplitude. We do not
detect any significant optical intra-night variations while the X-ray flux
changes significantly from orbit to orbit. We performed a $\chi^2$ test
against the hypothesis that the optical flux was constant. In order to
increase the signal to noise ratio we used light curves of the mean
optical magnitude per night. We find that the small amplitude variations
on time scales larger than a day are significant at $>99.9\%$ confidence
in all bands.

In order to quantify the amplitude variations we computed the fractional
excess variance, $\sigma^{2}_{rms}$ (Nandra et al. 1997). The square of
this parameter ($\sigma_{rms}$) gives the fractional amplitude of the
observed variability. In order to calculate the optical $\sigma_{rms}$ we
computed the mean de-reddened magnitudes per night, converted them into
flux and corrected for the contribution of the host galaxy to the mean
flux. For that reason, we used the PSF fitting routines of DAOPHOT in IRAF
to subtract the nucleus image from each frame and calculated the magnitude
before and after the nucleus subtraction. After this correction we found
that: $\sigma_{rms}(B)=2.1\pm 1.5\%, \sigma_{rms}(V)=1.9\pm 0.6\%,
\sigma_{rms}(R)=0.8\pm 1.5\%$ and $\sigma_{rms}(I)= 1.7 \pm 0.6\%$. The
average optical continuum $\sigma_{rms}$ is $1.6\pm 0.4\%$.

In the X-ray band, we combined SIS1/0 and GIS2/3 data and accumulated
light curves for the SIS $0.7-2$ keV, $2-5$ keV and the GIS $5-10$ keV
bands (we used the GIS data in that band because the GIS light curve has a
higher than the SIS light curve count rate). The estimation of
$\sigma_{rms}$ depends on the bin size and length of the light curve. The
use of light curves of the mean optical magnitude per night in the
estimation of the optical $\sigma_{rms}$ means that the optical data were
binned in $\sim 4$ hrs (the typical period during which the optical
observations were made each night). Therefore, the X-ray light curves were
accumulated in 16920 sec ($\sim 3$ orbital periods $\sim 4.7$ hours) and
were divided in two parts, so that each one of them has a length similar
to that of the optical light curves. We computed the $\sigma_{rms}$ for
each part and took their mean as our final estimate. Our results are as
follows:
$\sigma_{rms}(0.7-2$keV)$=29.5\pm1.5\%,\sigma_{rms}(2-5$keV)$=28.5\pm1.5\%$,
and $\sigma_{rms}(5-10$keV$)=27\pm2\%$.

\subsection{Cross Correlation analysis}
  
In order to examine the cross-links between variations in different bands
we estimated the cross-correlation functions using the Z-transformed
Discrete Correlation Function (ZDCF) of Alexander (1997) and the Discrete
Correlation Function (DCF) of Edelson \& Krolik (1988). Quoted errors on
the ZDCF lags correspond to $68\%$ confidense limits and were estimated
using the Monte Carlo methods described by Peterson et al. (1998). All the
X-ray light curves were binned in 5640 sec while for the optical band we
used the light curves of the mean flux per night.

\begin{figure}
\psfig{figure=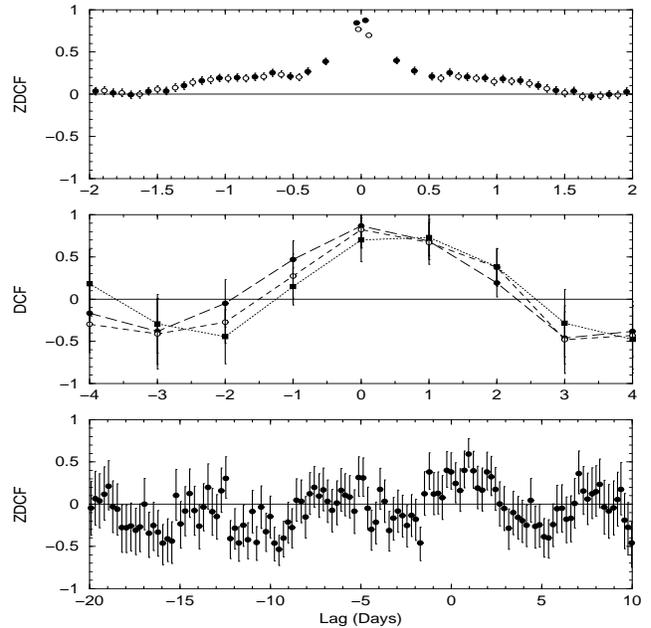,height=8.5truecm,width=8.5truecm,angle=0,%
  bbllx=5pt,bblly=45pt,bburx=550pt,bbury=715pt}
\caption[]{Cross correlation functions between: i) the $0.7-2/ 2-5$ keV
and $0.7-2/5-10$ keV X-ray bands (top plot), ii) the $B/V$, $B/R$ and
$B/I$ optical light curves (filled circles, open circles and filled boxes
respectively; middle plot) and iii) the $0.7-5$ keV/$B$ band light curves
(bottom plot). In all plots a positive lag means that the first listed
light curve leads the second. Note that the correlation function was
computed using the ZDCF technique for the upper and bottom plot, and the
DCF technique for the middle plot (see text).} 
\end{figure}

Fig. 3 (upper plot) shows the ZDCF between the SIS $0.7-2$ keV and the SIS
$2-5$ keV, GIS $5-10$ keV light curves (filled, open circles
respectively). The highest points are at lags $\tau=0.03^{+0.23}_{-0.29}$
days and $\tau=-0.02^{+0.28}_{-0.24}$ days respectively. The errors
associated with the maximum peak are quite large, probably because there
are a few points arount the peaks in the ZDCF, and the use of a large bin
size for the light curves. Therefore, we computed the cross correlation
between the same bands using 128 sec binned light curves. The maximum peak
was at lag zero. We conclude that there are no measurable delays between
the X-ray bands. If there exist any delays between the X-rays bands they
have to be smaller than $\sim$ few tens of seconds.

In the same Fig. (middle plot) we plot the cross correlation between the
$B/V$, $B/R$ and $B/I$ light curves (filled circles connected with a
long-dashed line, open circles connected with a dashed line and filled
squares connected with a dotted line respectively). Due to the small
number of points in the light curves the ZDCF (which bins data by equal
population) provides two estimates at $\sim$ zero lag and then at lags $>
2$ days. Since we are interested in the estimation of the correlation
coefficient at a more evenly spaced grid of lags, we used instead the DCF
technique which bins data in time bins of equal width. We used a lag bin
size of $1$ day and kept the DCF estimates at those lags with at least 10
individual points contributing to the DCF. Fig.~3 shows that the optical
variations are well correlated across the different bands (the DCF peak is
larger than 0.7 in all cases). The $V$, $R$ and $I$ bands appear to be
delayed with respect to the $B$ band light curve with the delay increasing
towards longer wavelengths. A parabolic fit is centered at $\tau_{B-V} =
0.2, \tau_{B-R}=0.45$ and $\tau_{B-I}=0.65$ days. The small number of
points in the light curves does not allow us to compute accurate errors on
these lag measurements. As a conservative estimate on the lag uncertainty
we accept the $(-1$ day,$+2$ days) interval beyond which the correlation
is $\sim 0$. We conclude that the small amplitude optical variations vary
nearly simultaneously at all bands with at most a $\sim +2$ day lag
between the $B$ and the $I$ band.

Finally we computed the ZDCF between the X-rays and the $B$ band light
curve (lower plot in Fig.~3). In the X-rays, in order to increase the
signal to noise we combined SIS1/0 data and accumulated a light curve in
the $0.5-5$ keV band (the addition of the $5-10$ keV data, even if we
combine GIS2/3, adds more to the noise than the signal, because of the
steep X-ray energy spectrum of the source).  The ZDCF is plotted at lags
where at least half of the optical light curve points contribute to the
estimation of the correlation (this is the reason that the center of the
plot is not at zero lag). The ZDCF peak is $r_{max}=+0.59$ at lag
$\tau=0.8^{+0.7}_{-1.1}$ days. Similar results are obtained when we cross
correlate the X-ray light curve with the other optical light curves as
well. To assess quantitatively the significance of the correlation, we
have carried out Monte Carlo simulations as follows. Synthetic optical
light curves with an $ \alpha = -1, -1.5$ and $-2$ red noise power
spectral density functions were created following the method of Timer and
K\"onig (1995).  The simulated light curves had the same variance and
sampling pattern as the real optical light curve. Random Gaussian
``errors", equal to the average error in the real light curve were also
added to the points in the simulated light curves. The ZDCF between the
synthetic optical and the real X-ray light curve was then computed, and
the maximum peak between lags $-20$ and $10$ days was kept. The whole
process was repeated 2000 times for each choice of optical power spectrum.
In this way the sample distribution of the ZDCF peaks can be computed. We
found that the observed correlation is significant at the 87$\%$, $85\%$
and $80\%$ level for a $-1, -1.5$ and $-2$ optical power spectrum,
respectively. We conclude that the maximum correlation coefficient in the
X-ray/optical ZDCF is statistically not significant in the sense that the
observed correlation could be due to the red noise character of the
variations in the light curves.

\section{Discussion}

We have presented a temporal analysis of a $\sim 30$day long ASCA
observation of \ark together with simultaneous optical monitoring
observations. Strong variability is observed in all X-ray bands. Optical
variations are also observed. Although statisctically significant, their
amplitude is much smaller than the amplitude of the X-ray variations. The
variations within the optical bands are strongly correlated. The B/V,R,I
cross-correlation functions show a sharp peak at a lag which increases
from the $B/V$ to the $B/I$ cross-correlation. However, the increase of
the delay within the optical bands is not statistically significant and
the cross-correlation results are consistent with the hyposthesis that the
optical variations vary almost simultaneously at all bands. We also
observe a peak at $\sim$ zero lag in the X-ray/optical cross-correlation
function. Its presence implies that the optical and X-ray variations are
positevely correlated but the statistical significance of the peak is low.

We now discuss our results in terms of models for AGN which involve X-ray
reprocessing mechanisms for the production of the UV/optical emission in
these objects. Due to the low statistical significance of the
X-ray/optical cross-correlation peak and of the time delays within the
optical bands, the scenarios that we discuss below based on these results
should be considered as suggestive physical interpretations.  Denser
optical light curves are needed in order to confirm the correlation
between the X-ray and optical variations and the existence of time delays
within the optical band in this source, allowing us a more quantitatively
discussion of the predictions of the respective physical models.

It is generally assumed that the X-rays in AGN arise from within a few
innermost radii ($\sim 3R_s$, where $R_{s}$ is the Schwarzschild radius)
of a supermassive black hole. There is lot of evidence from X-ray energy
spectral observations, mainly the 6.4 keV iron line and the so called
``Compton bump" (Nandra \& Pounds 1994), that the X-ray source in AGN
illuminates a relatively dense and cool material (i.e. the accretion
disc).  If that is the case, the heated disc is expected to radiate in the
ultraviolet/optical as well. So X-rays could be responsible for some or
all of the optical radiation via reprocessing (Guilbert \& Rees 1988). In
fact, some Comptonisation models which provide a good agreement with the
observed X-ray energy spectra of AGN, assume a specific geometry which
involves a slab of neutral material subtending a solid angle of $2\pi$ sr
to the X-ray source located above the slab (eg Haardt \& Maraschi 1991).
The X-ray luminosity of \ark is $L_{0.5-10~{\rm keV}} \sim 2 \times
10^{44}$ erg~s$^{-1}$ (Leighly 1999a, Vaughan et al. 1999). The sum of the
mean $B,V,R$ and $I$ luminosity during our monitoring campaign is $\sim
0.9\times 10^{44}$ erg~s$^{-1}$. Under the assumption that half of the the
X-ray flux is heating the disc, we expect the X-rays to affect the disc
output significantly in \arkp Consequently, we should observe optical
variations which follow those in the X-ray band, perhaps smoothed out
(below a certain time scale) due to geometric, light travel time and
variability amplitude of \ark is $\sim 20$ times that in the optical. This
result alone strongly suggests that most of the optical emission in \ark
is {\it not} reprocessed X-ray emission by optically thick material.
Either only a small part of the disc is seen by the X-ray source (i.e. the
X-ray source is {\it not} located above the disk but in its innermost part
and has a small height so the solid angle subtended by the disc is small)
or the X-rays are not isotropically emitted (i.e. most of the X-ray flux
is emitted away from the disc).

Furthermore, our optical campaign revealed small amplitude, but
statistically significant optical variations. These are observed at all
optical bands and are well correlated (with a maximum time lag of $\sim 2$
days) within the different bands. What could be their origin? One
possibility is that some part of the X-ray emission is, after all, heating
the disk, producing a reprocessed, variable optical component. In this
case, we expect to observe a time delay between the reprocessed optical
and X-ray emission of the order of $\tau \sim R/c$, where $R$ is the
distance between the X-ray source and the re-processing region. We do
observe a time lag of $\tau\sim 1$ day between X-ray and optical
variations. If this is real (which implies that the low significance of
the X-rays/optical correlation is due to the small number of points in the
optical light curves) we also expect to observe time delays within the
optical band. Assuming the temperature structure of a steady-state
accretion disc these delays should follow the relationship $\tau \propto
\lambda^{4/3}$ (Collier et al. 1998). In this case,
$\tau_{B-V}/\tau_{B-R}=(\lambda_{B}^{4/3}-\lambda_{V}^{4/3})/
(\lambda_{B}^{4/3}-\lambda_{R}^{4/3})=0.4$ and
$\tau_{B-V}/\tau_{B-I}=(\lambda_{B}^{4/3}-\lambda_{V}^{4/3})/
(\lambda_{B}^{4/3}-\lambda_{I}^{4/3})=0.22$. Interestingly, the ratio of
the respective observed lags (see section 3.2) is $\sim 0.5$ and $\sim
0.34$. This good agreement between the expected and the observed delays
between the optical bands supports the idea that the variable optical
component is reprocessed X-ray emission from optically thick material. The
$3\sigma$ upper limit on the time lag between X-rays and optical light
curves is $3.5$ days. Consequently, the optical emission region should be
at a distance smaller than $R \sim 10^{16}$cm from the black hole, i.e. at
$\sim 3\times 10^{4} - 3\times 10^3~R_s$ of a $10^{6} - 10^7$~M$_\odot$
black hole.

On the other hand, optical variations with delays between the optical
bands are also expected if there are instabilities propagating through an
accretion disc. In the case of a standard thin accretion disc with
sub-critical accretion rate the known mechanism for the instability
propagations (sound waves) is generally too slow to account for time lags
of the order of $\sim$ a few days (e.g. Krolik \etal 1991; Molendi,
Maraschi \& Stella 1992). If NLS1s have low black hole masses and
near-critical accretion rates the sound waves can propagate at a very high
speed, however detailed calculations in order to investigate whether the
sound crossing time in this case could explain the observed delays are
beyond the scope of the present work. Perhaps a more natural explanation
for the low-amplitude optical variations is that they are due to a flux
modulation of a region that does not dominate the optical emission. For
example, the innermost part of a standard accretion disk around a
supermassive black hole is expected to emit as a black body mainly at the
UV. Some of this radiation will be emitted in the optical band as well.
Consequently, if that region is variable due to a thermal or viscous
instability one expects simultaneous UV/optical variations and the
variability fractional amplitude to decrease with frequency. We find no
evidence of energy dependent variability amplitudes, although a longer
optical light curve is needed to reduce the errors on the variability
amplitudes and place stronger constraints. Densely sampled UV and optical
monitoring, simultaneously with X-ray, is needed in order to investigate
further the origin of the small amplitude optical variations in \arkp

Our results on the amplitude of \ark optical variations are consistent
with previous optical observations of other NLS1s like NGC~4051 (Done et
al. 1990, Peterson et al. 2000) and IRAS~13224-3809 (Young et al. 1999;
but note Miller et al 2000, who observe larger amplitude variations
on short time scales in this source). In general, the optical variations
in NLS1s are of smaller amplitude than those observed in the X-rays on
time scales of $\sim$ day. Although the number of objects that have been
studied so far is small, the results suggest that there is no significant
reprocessing of X-rays into the optical waveband in NLS1s. Optical
variability amplitudes significantly smaller than those in the X-rays have
also been observed in ``typical" Seyfert 1 galaxies like NGC~3516 and
NGC~7469 (Edelson et al. 2000, Collier et al. 1998). Perhaps then, more
complex geometrical and physical configurations have to be considered for
the Seyefrt galaxies, as a group, instead of the currently used simple
reprocessing scenarios. The similarity of the optical variability
amplitudes implies a common origin for the optical variable component in
Seyfert 1s. One possibility is reprocessing from optically thick material
of a small fraction of the X-ray flux. The NGC~469 data support this
hypothesis (Collier et al. 1998); the present data on \ark are also
consistent with it but this is not the case with NGC~3516 (Edelson et al.
2000). If on the other hand the optical variability is due to
instabilities of the accretion disc, the similarity of the optical
variability amplitudes is a rather unexpected result if NLS1s have black
hole masses smaller and accretion rates higher than other AGNs. The data
so far do not permit a detailed comparison between NLS1s and typical S1s;
intense optical observations of many objects from both classes is needed
to clarify this issue.

\vskip 0.4cm
\begin{acknowledgements}
We thank J. Englhauser and T. Reiprich and for helping us with the
observations.  Part of this work was done in the TMR research network
'Accretion onto black holes, compact stars and protostars' funded by the
European Commission under contract number ERBFMRX-CT98-0195.  Skinakas
Observatory is a collaborative project of the University of Crete, the
Foundation for Research and Technology-Hellas, and the
Max--Planck--Institut f\"ur extraterrestrische Physik.
\end{acknowledgements}

\end{document}